\PassOptionsToPackage{unicode}{hyperref}
\PassOptionsToPackage{hyphens}{url}
\PassOptionsToPackage{dvipsnames,svgnames,x11names}{xcolor}
\documentclass[
  12pt]{article}

\usepackage{amsmath,amssymb}
\usepackage{comment}
\usepackage{iftex}
\ifPDFTeX
  \usepackage[T1]{fontenc}
  \usepackage[utf8]{inputenc}
  \usepackage{textcomp} 
\else 
  \usepackage{unicode-math}
  \defaultfontfeatures{Scale=MatchLowercase}
  \defaultfontfeatures[\rmfamily]{Ligatures=TeX,Scale=1}
\fi
\usepackage{lmodern}
\ifPDFTeX\else  
\fi
\IfFileExists{upquote.sty}{\usepackage{upquote}}{}
\IfFileExists{microtype.sty}{
  \usepackage[]{microtype}
  \UseMicrotypeSet[protrusion]{basicmath} 
}{}
\makeatletter
\@ifundefined{KOMAClassName}{
  \IfFileExists{parskip.sty}{%
    \usepackage{parskip}
  }{
    \setlength{\parindent}{0pt}
    \setlength{\parskip}{6pt plus 2pt minus 1pt}}
}{
  \KOMAoptions{parskip=half}}
\makeatother
\usepackage{xcolor}
\makeatletter
\ifx\paragraph\undefined\else
  \let\oldparagraph\paragraph
  \renewcommand{\paragraph}{
    \@ifstar
      \xxxParagraphStar
      \xxxParagraphNoStar
  }
  \newcommand{\xxxParagraphStar}[1]{\oldparagraph*{#1}\mbox{}}
  \newcommand{\xxxParagraphNoStar}[1]{\oldparagraph{#1}\mbox{}}
\fi
\ifx\subparagraph\undefined\else
  \let\oldsubparagraph\subparagraph
  \renewcommand{\subparagraph}{
    \@ifstar
      \xxxSubParagraphStar
      \xxxSubParagraphNoStar
  }
  \newcommand{\xxxSubParagraphStar}[1]{\oldsubparagraph*{#1}\mbox{}}
  \newcommand{\xxxSubParagraphNoStar}[1]{\oldsubparagraph{#1}\mbox{}}
\fi
\makeatother

\usepackage{longtable,booktabs,array}
\usepackage{calc} 
\usepackage{etoolbox}
\makeatletter
\patchcmd\longtable{\par}{\if@noskipsec\mbox{}\fi\par}{}{}
\makeatother
\IfFileExists{footnotehyper.sty}{\usepackage{footnotehyper}}{\usepackage{footnote}}
\makesavenoteenv{longtable}
\usepackage{graphicx}
\makeatletter
\def\maxwidth{\ifdim\Gin@nat@width>\linewidth\linewidth\else\Gin@nat@width\fi}
\def\maxheight{\ifdim\Gin@nat@height>\textheight\textheight\else\Gin@nat@height\fi}
\makeatother
\setkeys{Gin}{width=\maxwidth,height=\maxheight,keepaspectratio}
\makeatletter
\def\fps@figure{htbp}
\makeatother

\makeatletter
\@ifpackageloaded{caption}{}{\usepackage{caption}}
\AtBeginDocument{%
\ifdefined\contentsname
  \renewcommand*\contentsname{Table of contents}
\else
  \newcommand\contentsname{Table of contents}
\fi
\ifdefined\listfigurename
  \renewcommand*\listfigurename{List of Figures}
\else
  \newcommand\listfigurename{List of Figures}
\fi
\ifdefined\listtablename
  \renewcommand*\listtablename{List of Tables}
\else
  \newcommand\listtablename{List of Tables}
\fi
\ifdefined\figurename
  \renewcommand*\figurename{Figure}
\else
  \newcommand\figurename{Figure}
\fi
\ifdefined\tablename
  \renewcommand*\tablename{Table}
\else
  \newcommand\tablename{Table}
\fi
}
\@ifpackageloaded{float}{}{\usepackage{float}}
\floatstyle{ruled}
\@ifundefined{c@chapter}{\newfloat{codelisting}{h}{lop}}{\newfloat{codelisting}{h}{lop}[chapter]}
\floatname{codelisting}{Listing}

\makeatother
\makeatletter
\@ifpackageloaded{caption}{}{\usepackage{caption}}
\@ifpackageloaded{subcaption}{}{\usepackage{subcaption}}
\makeatother

\ifLuaTeX
  \usepackage{selnolig}  
\fi
\usepackage[]{natbib}
\usepackage{bookmark}

\IfFileExists{xurl.sty}{\usepackage{xurl}}{} 
\hypersetup{
  pdftitle={Title},
  pdfauthor={Author 1; Author 2},
  pdfkeywords={3 to 6 keywords, that do not appear in the title},
  colorlinks=true,
  linkcolor={blue},
  filecolor={Maroon},
  citecolor={Blue},
  urlcolor={Blue},
  pdfcreator={LaTeX via pandoc}}

\newcommand{\anon}{1}

\usepackage{multirow}

\usepackage{verbatim}
\newcommand{\code}[1]{\texttt{#1}} 

\usepackage{tcolorbox}
\tcbuselibrary{listings, breakable}

\newtcblisting{codesnippet}{
  listing only,
  listing engine=listings,
  breakable,
  colback=lightgray!15,
  colframe=lightgray!15,
  top=0mm,
  bottom=0mm,
  listing options={
    basicstyle=\footnotesize\ttfamily,
    columns=fullflexible,
    keepspaces=true
  }
}

\begin{document}

\def\spacingset#1{\renewcommand{\baselinestretch}%
{#1}\small\normalsize} \spacingset{1}


\if1\anon
{
  \title{\bf Applying the Weibull Shape Parameter test for signal detection in pharmacovigilance using the R package WSPsignal}
  \author{Julia Dyck\\
    Biostatistics and Medical Biometry, Medical School OWL, Bielefeld University\\
    and \\
    Odile Sauzet \\
    School of Public Health, Bielefeld University \&\\ Department of Business Administration and Economics, Bielefeld University}
  \maketitle
} \fi

\if0\anon
{
  \bigskip
  \bigskip
  \bigskip
  \begin{center}
    {\LARGE\bf Title}
\end{center}
  \medskip
} \fi

\bigskip
\begin{abstract}
Post‑marketing pharmacovigilance relies on statistical signal detection methods to identify potential adverse drug reactions. The Weibull shape parameter (WSP) test concept exploits temporal information (electronic health records) to assess the hazard of an adverse event over time after drug initiation. A statistically significant deviation from constancy results in a signal. The WSP framework comprises a family of tests that differ with respect to the estimation approach (frequentist or Bayesian), the chosen time‑to‑event distribution (Weibull, double Weibull, power generalized Weibull) for hazard modeling, and test specification parameters. 

 To facilitate practical application and encourage consideration of the WSP signal detection test in future research, we developed the R package WSPsignal. The package consolidates all functionalities required for WSP testing into a unified, open‑source interface. It enables practitioners and researchers to apply default test specifications or perform simulation‑based tuning to identify the optimal test for a given data scenario.

We illustrate the package functionalities in two examples to follow along. In a large‑sample setting ($\approx 20\,000$ observations), a frequentist WSP test is considered. In a small-sample setting ($\approx 1\,000$ observations), a Bayesian WSP test is chosen. The additional test specifications are optimized through simulation‑based tuning.
\end{abstract}

\noindent%
\vfill

\newpage
\spacingset{1.8} 

\section{Introduction}\label{sec-intro}

Detecting potential adverse drug reactions (ADRs; \citealp{EMAwebsiteADR}) from routinely collected health data is a key element of post‑marketing pharmacovigilance. Signal detection methods aim to identify new or unexpected ADRs by testing for statistical associations between a prescribed drug and reported adverse events (AEs; \citealp{EMAwebsiteAE, meyboom1997pharmacovigilance}). A raised signal indicates the need for further investigation of the drug-AE pair under investigation. Because clinical trials before market approval typically involve limited and selected populations, some risks may only become apparent once a medicine is widely used. Detecting such signals is therefore crucial to maintain an up‑to‑date safety profile and can lead to label changes, usage restrictions, or market withdrawal \citep{hauben2005, noren2010temporal, sauzet2022}.

To perform signal detection based on longitudinal data, a range of statistical methods was developed: disproportionality approaches (originally conceptualized for spontaneous reports) adjusted for longitudinal data comparing rates of occurrence between a single drug and the rest of the dataset \citep{zorych2013disproportionality,schuemie2011lgps}, temporal pattern analysis \citep{noren2010temporal}, exposure models \citep{exposure_models2024preprint} and more (see for example the review by \citealp{coste2023methodreview}).

In this article, we focus on time-to-event-based signal detection tests that leverage temporal information in longitudinal data. These have been developed to evaluate whether the risk of an AE changes over time since drug prescription, forming a family of so-called Weibull shape parameter (WSP) tests \citep{cornelius2012, sauzet2022, dyck2025bpgw}.
The WSP test principle relies on one or more shape parameters of Weibull‑type distributions to assess deviations from a constant hazard (instant rate of event). In the case of a non-constant hazard, a signal is raised indicating a possible statistical temporal correlation between the drug and the , thereby marking the latter as a potential ADR. 

Several variants of the WSP test have been proposed, but, apart from the simplest version, their implementation has so far required substantial custom coding and scattered resources, limiting their accessibility to applied researchers.
We therefore present the newly developed R package WSPsignal \citep{WSPsignal}, which consolidates existing methods into a single open‑source interface offering a range of WSP test functionalities. It enables the application of default test specifications and the identification of the preferred test specification via simulation-based tuning for the signal detection task at hand.

This article provides a practical introduction to the package. After recalling the WSP test concept and components, we use synthetic data to illustrate the step‑by‑step application of the WSP signal detection tests, demonstrate test tuning, and discuss how the approach can be adapted to different data scenarios.

\section{Weibull shape parameter signal detection tests}\label{sec-methods}

We recall the statistical methods that constitute the family of WSP signal detection tests originally presented in \citet{cornelius2012, sauzet2022} and \citet{dyck2025bpgw}.

The WSP test concept is based on the assumption of temporal association between drug prescription and the occurrence of a drug-related AE. 
It is expected that the biological process leading the AE temporarily increases the risk of experiencing the AE. This assumption is formalized as a non-constant hazard function for the time-to-event (TTE) data, defined as the duration from the date of prescription to the occurrence of the AE (the instantaneous risk of AE).
A lack of a temporal dependence between drug prescription and AE is assumed to lead to a constant hazard function. Weibull-type models have the property that constant and non-constant hazards can be identified through values of one or more distribution parameters.

\begin{figure}[t]
\begin{subfigure}{0.5\textwidth}
\centering
\includegraphics[width=\linewidth, height=8cm]{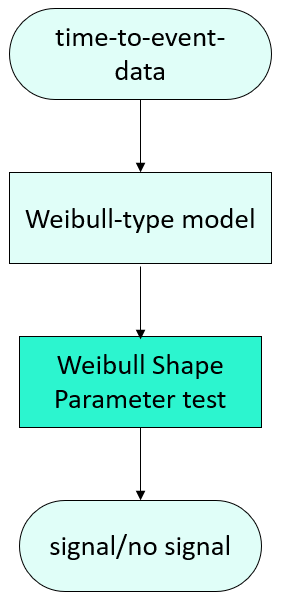} 
\end{subfigure}
\begin{subfigure}{0.5\textwidth}
\centering
\includegraphics[width=\linewidth, height=8cm]{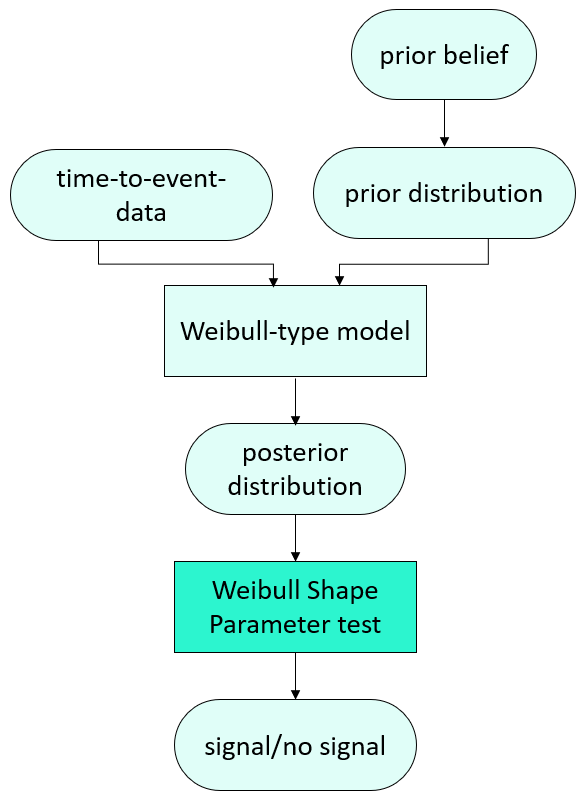}
\end{subfigure}
\caption{Workflow to get a signal/no signal from time-to-event data using a frequentist (left) or Bayesian (right) estimation approach.}
\label{fig:workflow}
\end{figure}

The WSP signal detection tests are designed for right-censored TTE data with time $t_i > 0$ since drug prescription and binary event status $d_i$ per observation $i$.
To get a signal for possible association, we first fit a Weibull-type model to the data and then perform a statistical hypothesis test on the shape parameters (see Figure \ref{fig:workflow}).
We present all available options for conducting frequentist and Bayesian WSP signal detection tests.

\subsection{Weibull-type models}\label{sec-methodmods}

Incorporated into the R package WSPsignal are the Weibull (W), the double Weibull (dW) (estimating two Weibull TTE models - one to the data $(t,d)$ as is and one to the data $(\tilde{t}, \tilde{d})$ censored at the middle of the observation period; \citealp{sauzet2022}), and the power generalized Weibull (pgW) model \citep{bagdonavicius2001, sauzet2022} with respective hazard functions

\begin{align}
    h_{W}(t) & = \nu \cdot \theta^{\nu}\cdot t^{\nu-1} \;\;,\\
    h_{dW,1}(t) & = \nu_1 \cdot \theta^{\nu_1}\cdot t^{\nu_1-1}\;  \text{ and }\; h_{dW,2}(\tilde{t}) = \nu_{2} \cdot \theta^{\nu_{2}}\cdot \tilde{t}^{\nu_{2}-1} \;\;,\\
    h_{pgW}(t) &  = \frac{\nu}{\gamma\theta^{\nu}}\cdot t^{\nu-1}\cdot  \left[ 1 + \left(\frac{t}{\theta}\right)^{\nu}\right]^{\frac{1}{\gamma-1}} \;\;,
\end{align}

scale parameter $\theta$, and shape parameters $\nu, \nu_1, \nu_2, \gamma$.
They can be estimated via frequentist maximum likelihood estimation \citep[p.81]{bagdonavicius2001}
or Bayesian posterior sampling \citep{dyck2025bpgw}. For the latter, prior distributions are to be specified for each model parameter (shape(s) and scale) by a distributional choice, prior mean, and prior standard deviation (SD). The available distributions include lognormal, gamma, fixed-scale lognormal, and fixed-scale gamma.

\subsection{Weibull shape parameter tests}

Signal detection results are derived from a significance test based on one or more shape parameters of hazard functions. The null and alternative hypotheses are summarized in Table \ref{tab:hypotheses} for all considered Weibull-type models.
\begin{table}
\caption{Null ($H_0$) and alternative ($H_1$) hypotheses for the Weibull shape parameter tests under frequentist and Bayesian Weibull (W), double Weibull (dW), and power generalized Weibull (pgW) models and sensitivity options 1, 2, and 3.
Bayesian test decisions are based on the posterior credibility interval and region of practical equivalence (ROPE) to the null value 1.
}
\label{tab:hypotheses}
\centering
\footnotesize
\begin{tabular}{|lll|}
    \hline
    model & \multicolumn{2}{l|}{null ($H_0$) and alternative ($H_1$) hypotheses}  \\ 
    \hline
    frequentist W & \multicolumn{2}{l|}{$\begin{array}{l}
        H_0: \nu = 1\\
        H_1: \nu \neq 1
    \end{array}$} \\ 
    
    frequentist dW & \multicolumn{2}{l|}{$\begin{array}{l}
        H_0: \nu_1 = 1 \text{ and } \nu_2 = 1\\
        H_1: \nu_1 \neq 1 \text{ or } \nu_2 \neq 1
    \end{array}$}\\
    
    frequentist pgW & \multicolumn{2}{l|}{$\begin{array}{l}
        H_0: \nu = 1 \text{ or } \gamma = 1\\
        H_1: \nu \neq 1 \text{ and } \gamma \neq 1
    \end{array}$}\\[6pt] 
    
    \multirow{2}{*}{Bayesian W} & \multirow{2}{*}{option 1} &  $H_0$: ROPE of $\nu$ accepted\\
    & & $H_1$: ROPE of $\nu$ rejected or indecision \\
    
    \multirow{2}{*}{Bayesian W} & \multirow{2}{*}{option 2 \& 3} &  $H_0$: ROPE of $\nu$ accepted or indecision\\
    & & $H_1$: ROPE of $\nu$ rejected \\
    
    \multirow{2}{*}{Bayesian dW \& pgW} & \multirow{2}{*}{option 1} &   $H_0$: at least one ROPE accepted and no ROPE rejected\\
    & & $H_1$: at least one ROPE rejected or two indecisions \\
    
    \multirow{2}{*}{Bayesian dW \& pgW} & \multirow{2}{*}{option 2} &   $H_0$: at least one ROPE accepted or two indecisions\\
    & & $H_1$: at least one ROPE rejected and no ROPE accepted \\   
    
    \multirow{2}{*}{Bayesian dW \& pgW} & \multirow{2}{*}{option 3} &   $H_0$: at least one ROPE not rejected\\
    & & $H_1$: both ROPEs rejected \\     
    \hline
\end{tabular}
\end{table}

In the frequentist context, test decisions are made at a particular significance level $\alpha$ (or confidence level $1-\alpha$). In the Weibull case (W or dW), a signal is raised if at least one shape parameter statistically differs from $1$. For the pgW model, a signal is raised if both shape parameters statistically differ from $1$, as this was found to perform better than the theoretical alternative hypothesis of only one \citep{Sauzet2024}.

\begin{figure}
    \centering
    \includegraphics[width=\linewidth]{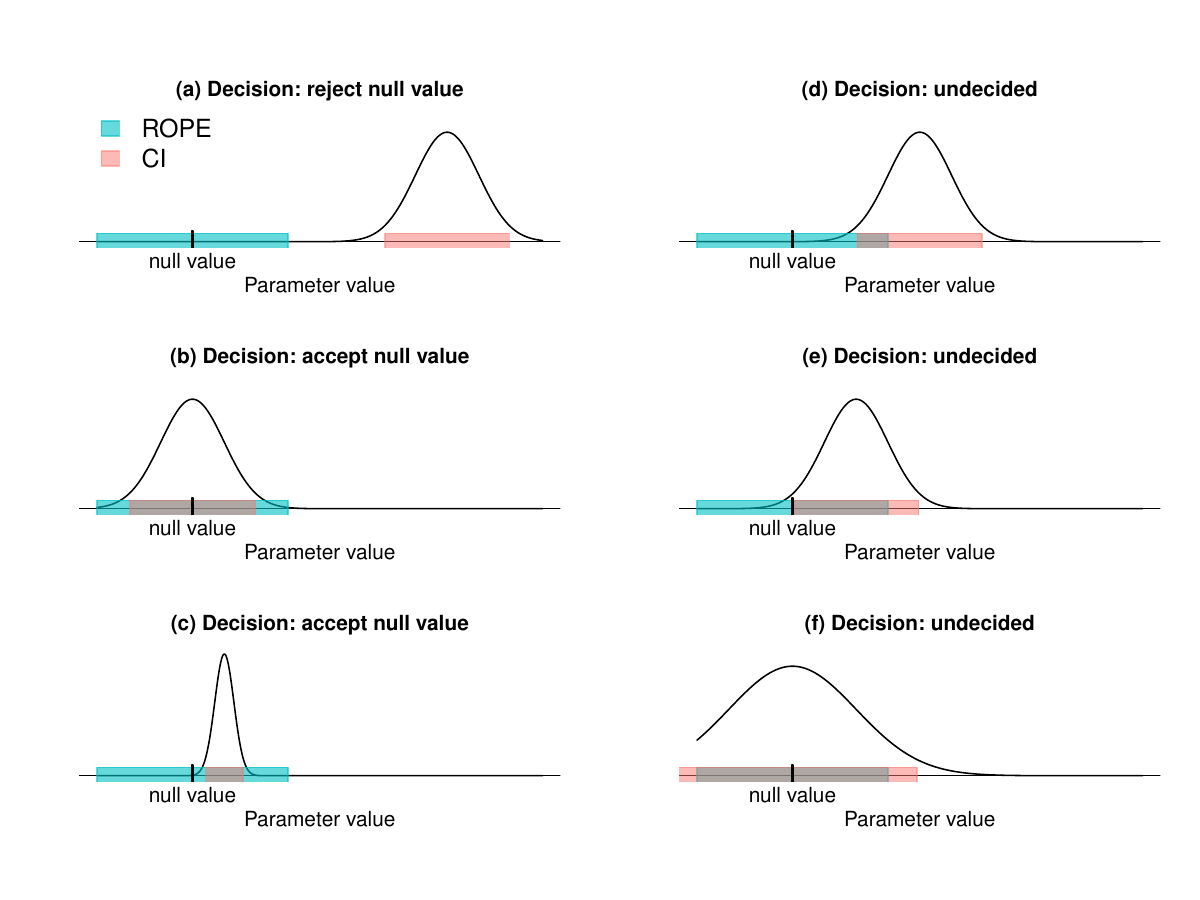}
    \caption{Decision rules for credibility interval (CI) + region of practical equivalence (ROPE) test based on graphic from \citet{kruschke2018}, Figure 1 and reprinted from \citet{dyck2025bpgw}, Figure 3.}
    \label{fig:CI+ROPE_outcomes}
\end{figure}

In the Bayesian context, the tests' results are based on the comparison between two intervals, namely the region of practical equivalence (ROPE) to the null value $1$ (parameter values within ROPE are deemed too close to 1 to statistically differ from it), and a posterior credibility interval (CI) at credibility level $1-\alpha$ obtained from the posterior sample (posterior parameter values within CI are deemed plausible given the posterior distribution; \citealp{kruschke2015, kruschke2018}). Each shape parameter's ROPE is either accepted (if CI lies completely in ROPE), rejected (if CI and ROPE are disjunct), or yields indecision (else) as an interim result (see Figure \ref{fig:CI+ROPE_outcomes}). Multiple sensitivity options (1 least to 3 most reserved; see Table \ref{tab:hypotheses}) are available to combine the interim results to a binary test output (signal/no signal).

Depending on the chosen test, the following arguments need specification:  
\begin{itemize}
    \item \textbf{Estimation approach}:
    \begin{itemize}
        \item Frequentist: 
            \begin{itemize}
                \item[$\rightarrow$] \textbf{TTE distribution} out of W, dW, pgW
                \item[$\rightarrow$] \textbf{confidence level} $1-\alpha \in (0,1)$
            \end{itemize}
        \item Bayesian:
            \begin{itemize}
                \item[$\rightarrow$] \textbf{TTE distribution} out of W, dW, pgW
                \item[$\rightarrow$] \textbf{prior mean} and \textbf{SD} for each model parameter
                \item[$\rightarrow$] \textbf{prior distribution} out of lognormal, gamma, fixed-scale-lognormal, fixed-scale-gamma
                \item[$\rightarrow$] \textbf{credibility level} $1-\alpha \in (0,1)$
                \item[$\rightarrow$] \textbf{posterior CI type} out of equal-tailed interval (ETI) or highest density interval (HDI; see \citep{kruschke2015} for definitions)
                \item[$\rightarrow$] \textbf{sensitivity option} for the combination of the shape parameters' interim test results.
            \end{itemize}
        \end{itemize}
\end{itemize}

We recommend choosing the estimation approach based on the context (sample size, prior knowledge). 
WSPsignal provides default test specifications for both estimation approaches as well as functions to identify the optimal model, and gives a study design and test setup for the application at hand by performing simulations and ranking the test specifications in terms of the area under the curve (AUC) of the receiver operating characteristic (ROC) graph with one threshold \citep{fawcett2004, lloyd1998}.

\section{Application of WSPsignal to two synthetic data cases
}\label{sec-appl1}
We demonstrate the WSP test procedure on the synthetic datasets \code{muscu} and \code{muscu2} embedded in WSPsignal. The \code{muscu} data are generated based on the descriptives of a real dataset from \citet{THINwebsite}, which was used in past case studies \citep{sauzet2013illustration, dyck2025bpgw} but is not openly available.
An additional dataset \code{muscu2} is generated with a smaller sample size. 
The test procedure is applied twice to illustrate the workflow using both the frequentist and Bayesian estimation approaches. 

The goal is to investigate whether the AE musculoskeletal pain should be flagged as a potential ADR associated with bisphosphonate treatment using WSP signal detection tests.
For this purpose, we 1. explore and describe the dataset, 2. apply a default WSP test (recommended based on past tunings), 3. perform a tuning to find an optimal WSP test considering the data characteristics at hand and apply the tuned test.

\subsection{Package installation}
The WSPsignal package can currently be installed and loaded with

\begin{codesnippet}
remotes::install_github(repo = "julia-dyck/WSPsignal")
library(WSPsignal)  
\end{codesnippet}

from Github. We load some additional packages for the procedure.
\begin{codesnippet}
library(dplyr)     # for data maneuvering
library(muhaz)     # for non-parametric hazard estimation
library(future)    # for parallelization of simulation study    
\end{codesnippet}

\subsection{Application I: frequentist Weibull shape parameter test in a large-sample setting}

\subsubsection{Data and descriptives}

The TTE dataset \code{muscu} contains $19\,777$ observations with time and event-status information, where the event is musculoskeletal pain. Observations are censored after 365 days for all subjects.
We calculate some descriptive statistics with the following commands.
\begin{codesnippet}
table(muscu$status) # no. of events vs. censored data
table(muscu$status)/nrow(muscu) # rate of observed AEs
muscu %>% filter(status == 1) %>% .$time %>% unlist() %>% mean() # mean TTE
muscu %>% filter(status == 1) %>% .$time %>% unlist() %>% sd() # sd of TTE
\end{codesnippet}
\noindent Within the observation period, 393 (2\%) of the $19\,777$ patients experienced musculoskeletal pain. The mean TTE is 170 days, with an SD of 80 days.

\begin{figure}[t]
    \centering
    \includegraphics[width=0.8\linewidth]{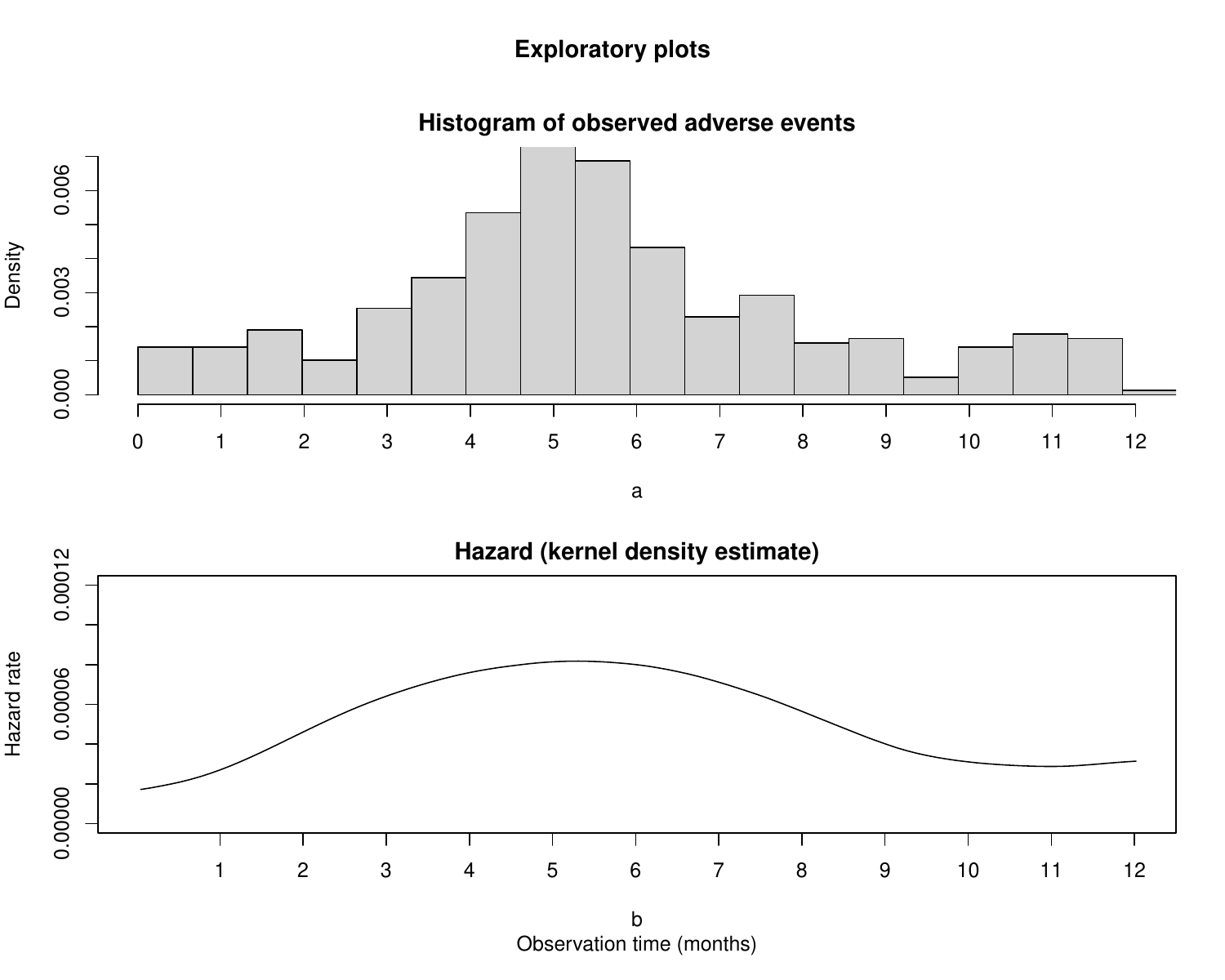}
    \caption{Graphics for exploratory analysis of the occurrence of musculoskeletal pain over one
year based on the synthetic dataset \code{muscu}.}
    \label{fig:muscu_exploratory}
\end{figure}
We gain a preliminary understanding of the risk of an AE over time by examining the distribution of observed events and a non-parametric hazard estimate.
\begin{codesnippet}
# histogram of event times
hist(muscu %>% filter(status == 1) %>% .$time, prob = T, xlim = c(0,365), 
    ylim=c(0,0.007), main = "Histogram of observed adverse events", 
     xlab = "Observation time (months)", breaks = 13, xaxt = "n")
axis(1,seq(0,365,30.4),0:12) # monthly tick marks instead of daily
# kernel density estimate for hazard
plot(muhaz(muscu$time, muscu$status, min.time=1, max.time=365,
           bw.smooth=500),
    main = "Hazard (kernel density estimate)",
    xlab ="Observation time (months)", ylab="Hazard rate", 
    xlim=c(0,365), ylim = c(0,12*10^(-5)), xaxt = "n")
axis(1,seq(30,365,30.4),1:12) # monthly tick marks instead of daily
\end{codesnippet}
The histogram (Figure \ref{fig:muscu_exploratory} a) shows a higher proportion of events around the fifth month. The same tendency is reflected by the non-parametric hazard estimate (Figure \ref{fig:muscu_exploratory} b), showing an increase at the beginning of the study period, followed by a decrease from the sixth month onwards.


\subsubsection{Default test}
The dataset is large, and we assume
to have no prior knowledge. We therefore focus on frequentist WSP (FWSP) tests.
\citet{Sauzet2024} recommended a combination of frequentist pgW and dW shape parameter test or only the dW shape parameter test with confidence level $0.97$ for sample sizes of about $20\, 000$, observation duration of one year, background rate of $0.01$ and potential ADR rate of $0.05$ (see \citet{Sauzet2024}; supplementary information).
Application of the dW test with

\begin{codesnippet}
mod = fwsp_model(dat = muscu, tte.dist = "dw") # model as basis for the FWSP test
Double Weibull parameter estimates in rweibull parametrization:
  parameter     estimate
1     scale 13892.364061
2     shape     1.073686
3   scale.c  1900.099157
4   shape.c     1.851722
testres = fwsp_test(mod.output = mod, cred.level = 0.97) # FWSP test
testres 
fwsp_dw_0.97 
           1 
\end{codesnippet}

leads to a signal as indicated by the output 1.

\subsubsection{Test tuning}\label{subsec-tuning1}

Using a recommended test specification is easy to apply and might be sufficient if the data scenario at hand (or a very similar one) has already been considered in existing simulation studies for WSP test tuning, as is the case for the FWSP test used. 
For optimal performance of the FWSP signal detection test, it is advised to tune it to determine the best TTE distribution and confidence level. This is done by performing simulations whose steps we now describe for the subset of FWSP tests.

\paragraph*{Setting up the simulation scenarios}

We consider data scenarios that reflect the characteristics of the data under investigation, such as the sample size, AE rate, and observation duration.

We simulate sample sizes of $20\,000$ observations, set the background rate to $0.01$, and the potential ADR rate to $0$, $0.5$, and $1$. Relative SD of event times (SD divided by length of study period) is set to $97/365 \approx 0.27$ and study period to $365$.

For TTE distributions, we choose dW and pgW.
For confidence levels we consider $0.5, 0.55, \dots, 0.85, 0.9, 0.91,\dots, 0.98, 0.99, 0.991,\dots, 0.999$. 

For simulation control arguments, that is, the number of repetitions per data scenario and modeling approach combination, the size of result table batches to be saved, and the working directory to save simulation results, default values (100 repetitions, batch size of 10, current working directory) are given. We create a folder to store all simulation-related files and set it as the working directory.

The simulation setup is prepared with

\begin{codesnippet}
pc_list = sim.setup_sim_pars(N = 20000, # data scenarios
                             br = 0.01, 
                             adr.rate = c(0, 0.5), 
                             adr.relsd = 0.27,
                             study.period = 365,
                             est.approach = "f", # estimation approach
                             tte.dist = c("dw", "pgw"), 
                             cred.level = c(seq(0.5,0.9, by = 0.05), # test spec.
                                            seq(0.91,0.99, by = 0.01), 
                                            seq(0.991, 0.999, by = 0.001)),
                             reps = 100, # simulation control arguments
                             batch.size = 10,
                             resultpath = "demo_application1")
\end{codesnippet}
providing a list of all information.
The \code{pc\_list} is used throughout the execution and evaluation process. We therefore save it with 
\begin{codesnippet}
 save(pc_list, file = paste0(pc_list$add$resultpath, "/pc_list.RData"))      
\end{codesnippet}

The simulation can be executed in sequence with \code{sim.run(pc\_list)} or in parallel with 
\begin{codesnippet}
future::plan(multisession, workers = future::availableCores())
sim.run_parallel(pc_list = pc_list)     
\end{codesnippet}

In case of an interruption, {\code{sim.run()}} or {\code{sim.run\_parallel()}} can be executed again and will pick up where they stopped.

After completion of the simulations, we merge the result batches with

\begin{codesnippet}
sim.merge_results(pc_list, save = T)   
\end{codesnippet}

and save them as \code{res\_f.RData} ready for evaluation of the WSP tests.

\paragraph*{Evaluation of tests and finding the best tuning}

The considered models and confidence levels are evaluated based on the accuracy of the test (AUC, false positive rate [FPR], true positive rate [TPR], false negative rate [FNR] and true negative rate [TNR]). A ranking of test specifications in terms of AUC is obtained with
\begin{codesnippet}
perf = eval.calc_perf(pc_list = pc_list)
rank = eval.rank_auc(perf, tte.dist.subset = c("dw", "pgw"))
rank$rank.tab
\end{codesnippet}
\begin{table}[t]
\caption{Five best frequentist Weibull shape parameter test specifications.}
\label{tab:ranking_fwsp}
\centering
\footnotesize
\begin{tabular}{|r|llllllll|}
\hline
rank & test & TTE & conf. & AUC & FPR & TPR & FNR & TNR \\
 & type & dist. & level & & & & & \\
\hline
1 & fwsp & dw & 0.97 & 0.815 & 0.040 & 0.671 & 0.329 & 0.960 \\
2 & fwsp & dw & 0.96 & 0.813 & 0.060 & 0.687 & 0.313 & 0.940 \\
3 & fwsp & dw & 0.95 & 0.810 & 0.080 & 0.700 & 0.300 & 0.920 \\
4 & fwsp & dw & 0.91 & 0.808 & 0.120 & 0.737 & 0.263 & 0.880 \\
5 & fwsp & dw & 0.93 & 0.807 & 0.105 & 0.720 & 0.280 & 0.895 \\
\hline
\end{tabular}
\end{table}
\begin{figure}[t]
    \centering
    \includegraphics[width=0.8\linewidth]{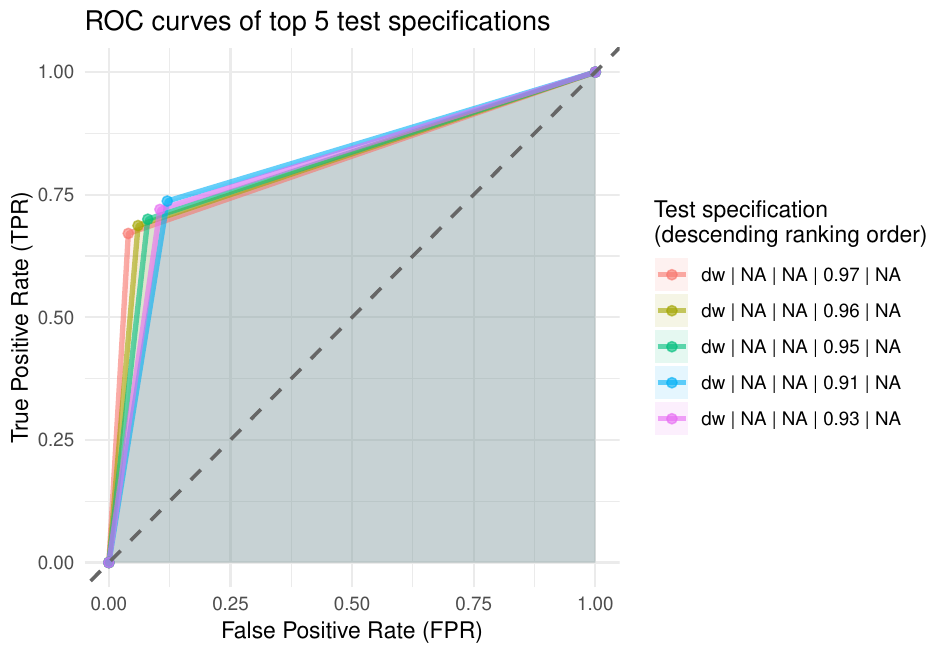}
    \caption {ROC curves representing the top five frequentist Weibull shape parameter test performances.}
    \label{fig:roc_curve_fwsp}
\end{figure}

leading to the results in Table \ref{tab:ranking_fwsp}. The ROC curves corresponding to the AUCs of the five best tests can be visualized with \code{eval.roc\_curve(rank\$rank.tab, n = 5)} (see Figure \ref{fig:roc_curve_fwsp}). The preferred FWSP test is based on the dW model with a $97\%$ confidence level.

Effects of simulation scenario parameters (sample size, background rate, ADR rate, prior belief, relative SD, and, in case of a Bayesian approach, distance of the prior belief to the truth) on the AUC of the best WSP test are also stored in the output of \code{eval.rank\_auc()}.

\begin{codesnippet}
rank$effect.of.N
      N   AUC   FPR   TPR   FNR   TNR
1 20000 0.815  0.04 0.671 0.329  0.96
rank$effect.of.br
     br   AUC   FPR   TPR   FNR   TNR
1  0.01 0.815  0.04 0.671 0.329  0.96
rank$effect.of.adr.rate
  adr.rate   AUC   FPR   TPR   FNR   TNR
1      0.5 0.794  0.04 0.628 0.372  0.96
2      1   0.837  0.04 0.713 0.287  0.96
rank$effect.of.adr.when
  adr.when   AUC   FPR   TPR   FNR   TNR
1     0.25 0.98   0.04 1     0      0.96
2     0.5  0.548  0.04 0.135 0.865  0.96
3     0.75 0.919  0.04 0.878 0.122  0.96
rank$effect.of.adr.relsd
  adr.relsd   AUC   FPR   TPR   FNR   TNR
1      0.27 0.815  0.04 0.671 0.329  0.96
rank$effect.of.dist.prior.to.truth
NULL
\end{codesnippet}
Here, the only simulation scenario parameters with variation are the ADR rate and expected TTE. 
According to \code{rank\$effect.of.adr.rate}, the test performs better when the ADR rate is higher.
The \code{rank\$effect.of.adr.when} shows that the test performs very well when the ADR occurs towards the beginning (AUC $= 0.98$) or end (AUC $= 0.92$) of the study period. 

As the FWSP test with tuned model (dW) and confidence level ($0.97$) specification equals the default FWSP test, its application to the \code{muscu} dataset leads to a signal as well.

\subsection{Application II: Bayesian Weibull shape parameter test in a small-sample setting}\label{sec-appl2}
To illustrate the tuning procedure when the data suggest using a Bayesian estimation approach, we generated a second synthetic dataset, \code{muscu2}, with a smaller sample size.

\subsubsection{Data and descriptives}
\begin{figure}[t]
    \centering
    \includegraphics[width=0.8\linewidth]{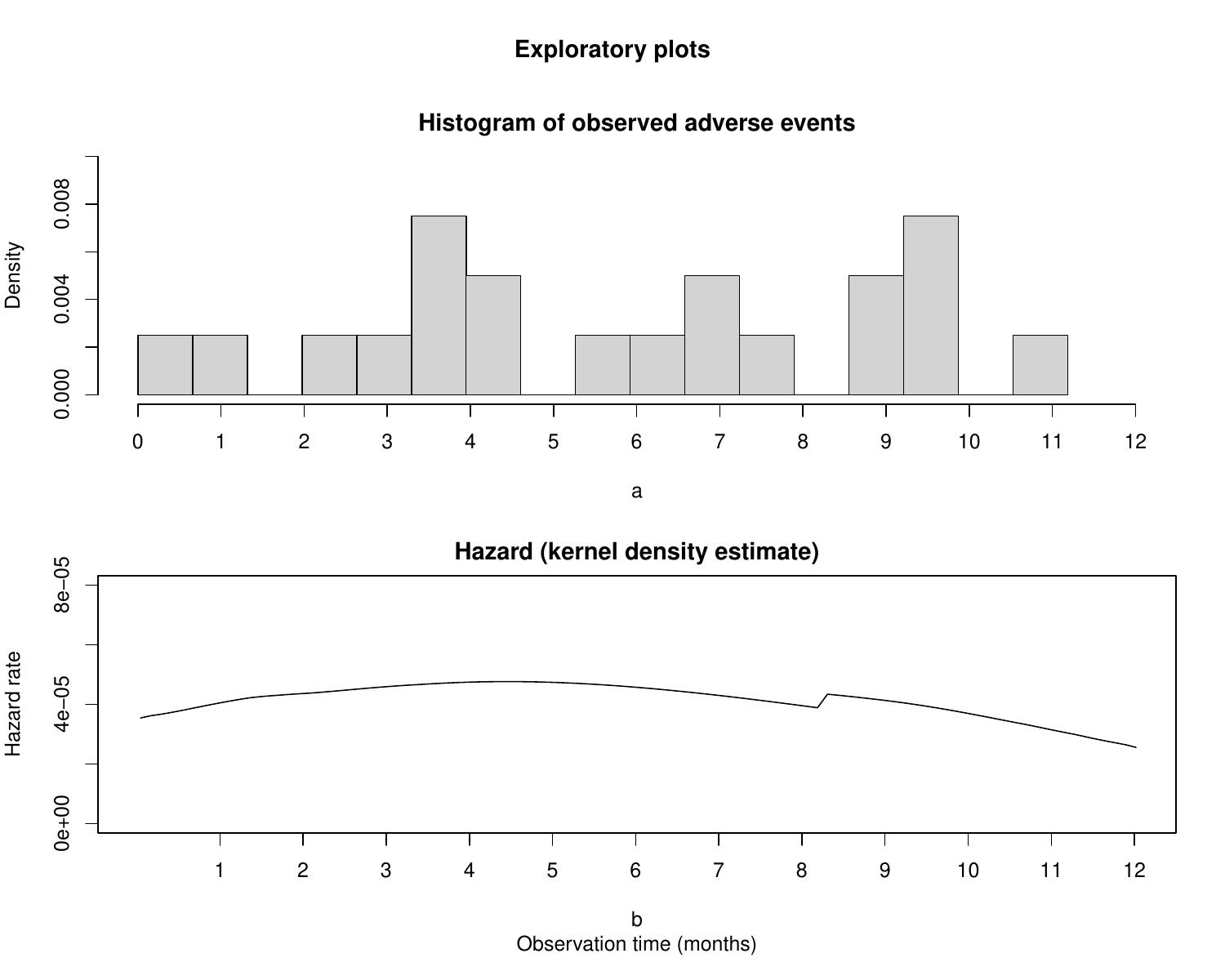}
    \caption{Graphics for exploratory analysis of the occurrence of musculoskeletal pain over one
year based on the synthetic dataset \code{muscu2}.}
    \label{fig:muscu2_exploratory}
\end{figure}

The sample size of the TTE dataset \code{muscu2} is $1\,208$. The data is censored after $365$ days. The event is observed for 20 (1.7\%) of the subjects. The mean TTE of the cohort is 176 days with an SD of 96 days. This is slightly indicated by the non-parametric hazard estimate showing a peak of risk after approximately five months (see Figure \ref{fig:muscu2_exploratory}).

\subsubsection{Default test}
\citet{dyck2025bpgw} recommended, based on simulations representing sample sizes of $500$ to $5\,000$, a background rate of $0.1$ and ADR rates of 0.5 or 1, the Bayesian WSP (BWSP) test with pgW model (only TTE distribution considered), lognormal prior distributions, an 80\% HDI for the posterior CI, and sensitivity option 2 for the combination rule.

As musculoskeletal pain is known to be an ADR of bisphosphonates that can occur at any time after intake \citep{adr_bis_muscu}, we set the prior means to correspond to a unimodal hazard with the expected TTE near the middle of the observation period.

Preparing the prior, fitting the model (Warning: this may take some time), and calling the test function with 

\begin{codesnippet}
muscu_pl_prior2 = tte2priordat(dat = muscu2, tte.dist = "pgw", prior.dist = "ll",
                              scale.mean = 20, scale.sd = 10, 
                              shape.mean = 5.5, shape.sd = 10, 
                              powershape.mean = 14, powershape.sd = 10)

mod2 = bwsp_model(muscu_pl_prior2) # fit model
Inference for Stan model: pgw_tte_lognormalprior.
4 chains, each with iter=11000; warmup=1000; thin=1; 
post-warmup draws per chain=10000, total post-warmup draws=40000.
         mean se_mean    sd n_eff Rhat
theta   60.16    0.19 26.68 19269    1
nu       1.26    0.00  0.36 18022    1
gamma  121.32    0.32 41.12 16443    1
lp__  -235.15    0.01  1.25 14496    1
Samples were drawn using NUTS(diag_e) at Thu Mar 26 16:59:35 2026.
For each parameter, n_eff is a crude measure of effective sample size,
and Rhat is the potential scale reduction factor on split chains (at 
convergence, Rhat=1).

bwsp_test(mod.output = mod2, # BWSP test
cred.level = 0.8, ci.type = "HDI", sensitivity.option = 2)
bwsp_pgw_ll_0.8_HDI_2 
                    1 
\end{codesnippet}

leads to a signal.

\subsubsection{Test tuning}

The recommended test specification presented in \citet{dyck2025bpgw} sufficiently represents the sample size at hand, but the AE rate and relative SD observed in the data are considerably lower than those in the simulations. We therefore conduct a new simulation study to determine which BWSP test specification is most appropriate, given the data characteristics at hand.
We tune the BWSP test with respect to the TTE distribution, credibility level, prior distribution, prior means, and SDs for the TTE distribution parameters, posterior CI types, and sensitivity options.

\paragraph*{Setting up the simulation scenarios}

Data scenarios considered should ideally be based on prior knowledge, the sample size, and observation duration available in the dataset.

We simulate samples with $1\, 000$ observations. The background rate is set to $0.01$. Potential ADR rates of $ 0$, $ 0.5$, and $1$ are considered. The relative SD of event times is set to $0.27$, and the study period to $365$.

We decide on the following TTE distributions: dW and pgW. For estimation, we choose the Bayesian approach, which requires specifying a prior distribution, prior mean, and SD for each parameter of the TTE distributions.

For prior distributions, we consider gamma and lognormal.
The next step is to provide vectors of prior means and SDs. Priors must be specified to reflect the belief that an ADR does not occur or occurs towards the beginning, middle, or end of the study period. It may be helpful to visualize the effect of chosen prior means on the hazard function to check whether the given form and expected time reflect the intended prior belief.

\begin{codesnippet}
# find prior means for double Weibull setting:
# uncensored Weibull parameters
plot_pgw(scale = 1, shape = 1, powershape = 1)     # "none" prior
plot_pgw(scale = 1, shape = 0.207, powershape = 1) # "beginning" prior
plot_pgw(scale = 180, shape = 1, powershape = 1)   # "middle" prior
plot_pgw(scale = 300, shape = 4, powershape = 1)   # "end" prior
# Weibull (censored) parameters
plot_pgw(scale = 1, shape = 1, powershape = 1)     # "none" prior
plot_pgw(scale = 1, shape = 0.207, powershape = 1) # "beginning" prior
plot_pgw(scale = 100, shape = 4, powershape = 1)   # "middle" prior
plot_pgw(scale = 1, shape = 1, powershape = 1)     # "end" prior

# find prior means for Power generalized Weibull parameters:
plot_pgw(scale = 1, shape = 1, powershape = 1)     # "none" prior
plot_pgw(scale = 20, shape = 5.5, powershape = 14) # "beginning" prior
plot_pgw(scale = 180, shape = 1, powershape = 1)   # "middle" prior
plot_pgw(scale = 300, shape = 4, powershape = 1)   # "end" prior
\end{codesnippet}

\begin{table}[t]
    \centering
    \footnotesize
    \caption{Mean of prior distribution under implemented prior beliefs about expected TTE.
    Prior standard deviation is set to 10 for all parameters in all prior belief cases.}
    \label{tab:sim_priorbelief}
    \begin{tabular}{|l|r|c|c|}
        \hline
        prior belief
        & a priori expected
        & \multicolumn{2}{c|}{a priori parameter means under} \\ 
        & TTE $E[t]$ 
        & dW model 
        & pgW model \\
        & 
        & (\verb|scale|, \verb|shape|, \verb|scale_c|, \verb|shape_c|)
        & (\verb|scale|, \verb|shape|, \verb|powershape|) \\
        \hline
        none
        & - 
        & $(1, 1, 1, 1)$
        & $(1, 1, 1)$ \\
        beginning
        & 91 days
        & $(1, 0.207, 1, 0.207)$
        & $(1, 0.207, 1)$ \\
        middle
        & 183 days
        & $(180, 1, 100, 4)$
        & $(20, 5.5, 14)$ \\
        end
        & 274 days
        & $(300, 4, 1, 1)$
        & $(300, 4, 1)$ \\
        \hline
    \end{tabular}
\end{table}

 \begin{figure}
    \centering
    \includegraphics[width = \textwidth]{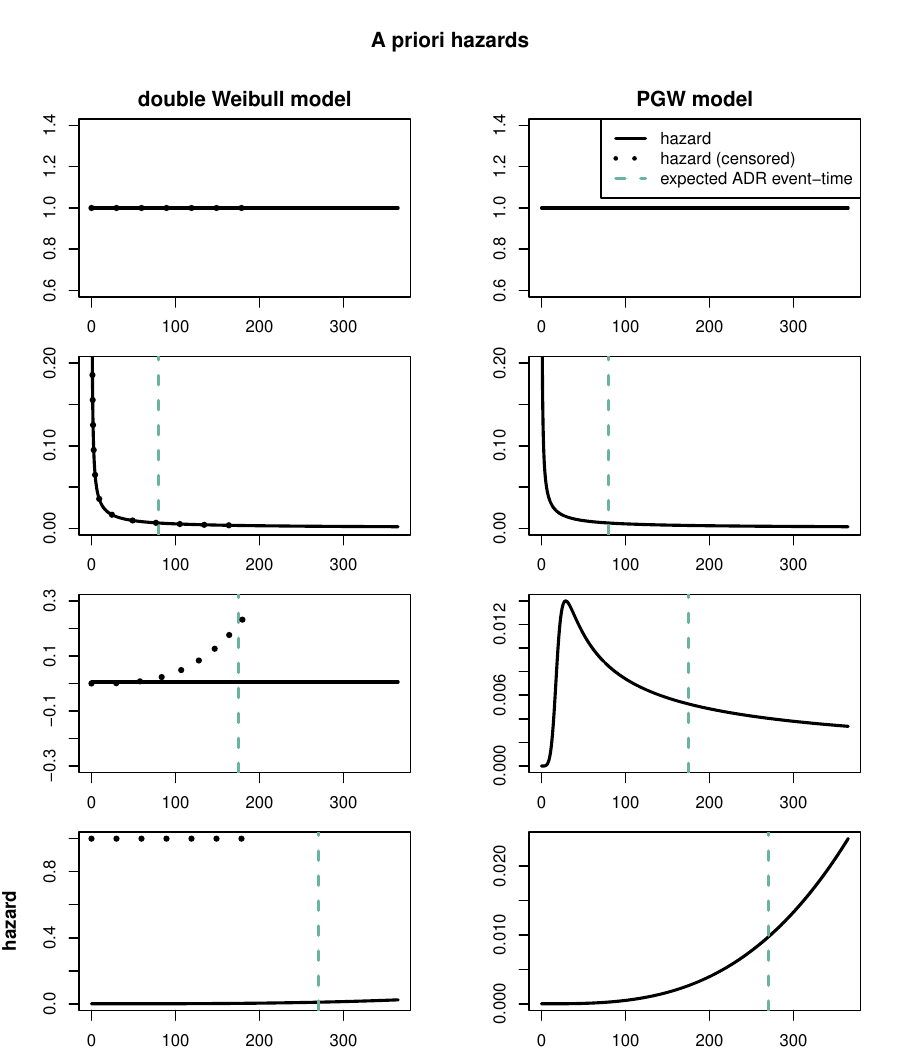}
    \caption{Hazard functions obtained under the prior belief that the adverse event is not an adverse drug reaction, or that the expected time-to-event is at the first, second, or third quarter of the study period (top to bottom) under a dW (left) or a pgW (right) model.}
    \label{fig:priorhazards}
\end{figure}
Possible prior mean vectors are provided in Table \ref{tab:sim_priorbelief}. Each parameter's prior SD is set to 10. Hazard functions obtained under the prior means are visualized in Figure \ref{fig:priorhazards}. 
For instance, the prior setup under the dW distribution reflecting the end belief (see Figure \ref{fig:priorhazards} bottom, left) is a combination of an increasing Weibull hazard for the whole study period and a constant hazard for the study period censored at the middle without fine-tuning the baseline hazard (prior scale mean), just roughly considering the trend.

We set up a list of all the prior parameters using \code{sim.prior\_template()}, as shown in the code below.

\begin{codesnippet}
fp_list = sim.priors_template(tte.dist = c("dw", "pgw"), prior.sds = 10) 

fp_list$dw[,2] = c(1, 1, 180, 300)   # scale prior means
fp_list$dw[,3] = c(1, 0.207, 1, 4)   # shape prior means
fp_list$dw[,4] = c(1, 1, 100, 1)     # scale_c prior means
fp_list$dw[,5] = c(1, 0.207, 4, 1)   # shape_c prior means

fp_list$pgw[,2] = c(1, 1, 20, 300)    # scale prior means
fp_list$pgw[,3] = c(1, 0.207, 5.5, 4) # shape prior means
fp_list$pgw[,4] = c(1, 1, 14, 1)      # powershape prior means
\end{codesnippet}

Next, we specify possible credibility levels, CI types reflecting the posterior distribution, and sensitivity options for combining the single-parameter test results. 
We consider $0.5, 0.55, \dots, 0.85, 0.9, 0.91,\dots, 0.98, 0.99, 0.991,\dots, 0.999$ for credibility levels, both CI types ETI and HDI, and all three sensitivity options. 

The simulation control arguments are set in the same manner as in Section \ref{subsec-tuning1}.
The number of repetitions per data-scenario and modeling-approach combination, and the size of result table batches to be saved, are set to default values (100 repetitions, batch size of 10). A new folder is created for all simulation-related files, and set as the result path.

The control arguments for the Bayesian sampling, namely the number of chains, the number of iterations, and the number of warmup iterations not included in the posterior samples, will be passed to the inside function \code{bwsp\_model()}. We stick to the default values of four chains with $11 000$ iterations and $1000$ warmup iterations each.

We prepare the simulation setup with

\begin{codesnippet}
pc_list2 = sim.setup_sim_pars(N = 1000, # data scenarios
                             br = 0.01, 
                             adr.rate = c(0, 0.5, 1), 
                             adr.relsd = 0.27,
                             study.period = 365,
                             est.approach = "b", # estimation approach
                             tte.dist = c("dw", "pgw"),
                             prior.dist = c("ll", "gg"),
                             fitpars.list = fp_list,
                             post.ci.type = c("ETI", "HDI"), # test specifications
                             cred.level = c(seq(0.5,0.9, by = 0.05), 
                                            seq(0.91,0.99, by = 0.01), 
                                            seq(0.991, 0.999, by = 0.001)), 
                             sensitivity.option = 1:3,
                             reps = 100, # simulation control arguments
                             batch.size = 10,
                             resultpath = "demo_application2",
                             stanmod.chains = 4, # Bayesian sampling control arguments
                             stanmod.iter = 11000,
                             stanmod.warmup = 1000
)
\end{codesnippet}
which returns a list of all the required information, and save it with 
\begin{codesnippet}
 save(pc_list2, file = paste0(pc_list$add$resultpath, "/pc_list2.RData"))      
\end{codesnippet}

The simulation is executed, and interim results are merged and saved as \code{res\_b.RData} in the same way as in Section \ref{subsec-tuning1}
whereby we advise parallelization, as repeated Bayesian sampling is costly in time.




\paragraph*{Evaluation of tests and finding the best tuning}

Under the Bayesian estimation approach, the first step of evaluation is to assess the computational performance of the Bayesian tests. Performance is assessed in terms of convergence issues, execution times, and effective sample sizes. This information is provided in form of tables and figures (see Appendix \ref{app:computational_perf}) by calling

\begin{codesnippet}
nonconv = eval.non_conv_cases(pc_list2, group.by = c("tte.dist", "prior.dist")
  tte.dist prior.dist total.planned total.notrun prop.notrun
1 dw       gg                  2800           27     0.00964
2 dw       ll                  2800           25     0.00893
3 pgw      gg                  2800            0     0      
4 pgw      ll                  2800            0     0

extimes = eval.execution_times(pc_list2, group.by = c("tte.dist", "prior.dist")
  group      min first_qu median  mean third_qu   max
1 dw - gg  1.98      2.49   2.63  2.71     2.90  3.72
2 pgw - gg 1.37      2.29   2.92  3.00     3.69  7.19
3 dw - ll  1.33      1.69   2.32  2.32     2.80  3.18
4 pgw - ll 0.896     1.69   2.66  2.94     3.73  9.70

neff = eval.eff_sample_sizes(pc_list2, group.by = c("tte.dist", "prior.dist", 
                             threshold = 10000)
  tte.dist prior.dist parameter prob.above.thr
1 dw       gg         shape1             0.743
2 dw       gg         shape2             0.934
3 dw       ll         shape1             0.499
4 dw       ll         shape2             0.251
5 pgw      gg         shape1             0.980
6 pgw      gg         shape2             0.958
7 pgw      ll         shape1             0.742
8 pgw      ll         shape2             0.675
\end{codesnippet}
Cases tending to a higher rate of non-convergence, higher execution time, and low effective sample size should not be considered in the final analysis. 

The pgW models show no convergence issues, whereas the dW models lead to convergence issues in $1\%$ of simulations under a gamma prior and in $0.9\%$ under a lognormal prior (Table \ref{tab:demoappl-musku2-notrun}). 
Execution times remain below 10 minutes across all model and prior-distribution choices (Figure \ref{fig:demoappl-musku2-runningtimes}). Proportions of effective sample sizes are highest for pgW models with gamma prior ($98\%$ for shape, $96\%$ for powershape), followed by dW models with gamma prior ($74\%$ for shape, $93\%$ for censored model shape)
and pgW models with lognormal prior ($74\%$ for shape, $68\%$ for powershape). The lowest proportions occur under dW models with a lognormal prior ($50\%$ for shape, $25\%$ for the censored model shape; see Figure \ref{fig:demoappl-musku2-effectivesamplesize}).
We conclude that the dW model with a lognormal prior should be excluded from the final analysis. Should a pgW model with lognormal prior be chosen for the final test, we need to pay attention whether the effective sample size exceeds the threshold of $10\,000$. If not, we need to increase the number of iterations for the Bayesian sampling.

In the second step of evaluation, we compare the BWSP test specifications along the performance measures AUC, FPR, TPR, FNR and TNR. All test specifications are ranked in terms of AUC with
\begin{codesnippet}
perf = eval.calc_perf(pc_list = pc_list2)
rank = eval.rank_auc(perf)
rank$rank.tab
\end{codesnippet}
\begin{table}[t]
\centering
\footnotesize
\begin{tabular}{|r|llllllllllll|}
\hline
rank & test & TTE & prior & prior & post. & cred. & sensitivity & AUC & FPR & TPR & FNR & TNR \\
 & type & dist. & dist. & SD & CI type & level & option & & & & & \\
\hline
1 & bwsp & pgw & ll & 10 & HDI & 0.60 & 3 & 0.655 & 0.397 & 0.707 & 0.293 & 0.603 \\
2 & bwsp & pgw & ll & 10 & HDI & 0.55 & 3 & 0.643 & 0.587 & 0.873 & 0.127 & 0.413 \\
3 & bwsp & pgw & ll & 10 & HDI & 0.65 & 3 & 0.643 & 0.227 & 0.513 & 0.487 & 0.773 \\
4 & bwsp & pgw & gg & 10 & HDI & 0.90 & 3 & 0.626 & 0.643 & 0.895 & 0.105 & 0.357 \\
5 & bwsp & pgw & gg & 10 & ETI & 0.90 & 3 & 0.620 & 0.700 & 0.940 & 0.060 & 0.300 \\
\hline
\end{tabular}
\caption{Five best Bayesian Weibull shape parameter test specifications.}
\label{tab:ranking_bwsp}
\end{table}

\begin{figure}
    \centering
    \includegraphics[width=0.8\linewidth]{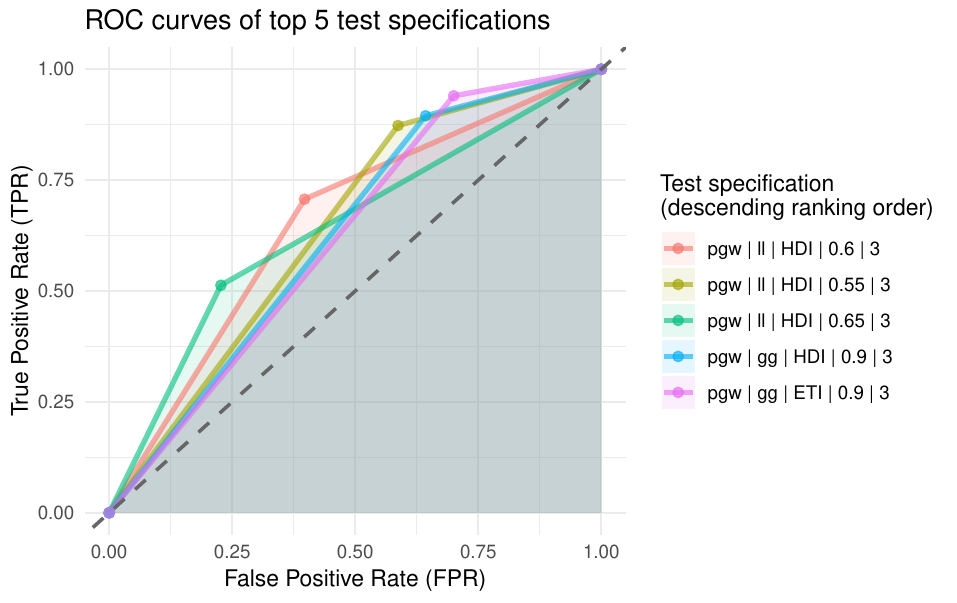}
    \caption{ROC curves representing the top five Bayesian Weibull shape parameter test performances.}
    \label{fig:roc_curve_bwsp}
\end{figure}

providing the ranking shown in Table \ref{tab:ranking_bwsp}. Again, we visualize the ROC curves corresponding to the AUCs of the five best tests with \code{eval.roc\_curve(rank\$rank.tab, n = 5)} (see Figure \ref{fig:roc_curve_bwsp}). 

The effects of simulation scenario parameters (sample size, background rate, ADR rate, prior belief, relative SD, and in case of a Bayesian approach, distance of the prior belief to the truth) on the AUC are also obtained with \code{eval.rank\_auc()}
\begin{codesnippet}
rank$effect.of.N
      N   AUC   FPR   TPR   FNR   TNR
1  1000 0.655 0.397 0.707 0.293 0.603
rank$effect.of.br
     br   AUC   FPR   TPR   FNR   TNR
1  0.01 0.655 0.397 0.707 0.293 0.603
rank$effect.of.adr.rate
  adr.rate   AUC   FPR   TPR   FNR   TNR
1      0.5 0.637 0.397 0.67  0.33  0.603
2      1   0.673 0.397 0.743 0.257 0.603
rank$effect.of.adr.when
  adr.when   AUC   FPR   TPR   FNR   TNR
1     0.25 0.6    0.19 0.39  0.61   0.81
2     0.5  0.792  0.16 0.745 0.255  0.84
3     0.75 0.572  0.84 0.985 0.015  0.16
rank$effect.of.adr.relsd
  adr.relsd   AUC   FPR   TPR   FNR   TNR
1      0.27 0.655 0.397 0.707 0.293 0.603
rank$effect.of.dist.prior.to.truth
  dist.prior.to.truth     AUC   FPR   TPR   FNR   TNR
1 correct specification 0.655 0.397 0.707 0.293 0.603
2 no ADR assumed        0.632 0.2   0.463 0.537 0.8  
3 one quarter off       0.667 0.338 0.671 0.329 0.662
4 two quarters off      0.451 0.515 0.418 0.582 0.485
\end{codesnippet}
 show that the test tends to work better when the ADR occurs towards the middle of the study period. Moreover, it highlights the importance of including prior knowledge only when its correctness is certain.

Applying the tuned BWSP (pgW model, lognormal prior distribution, HDI, credibility level 0.6, sensitivity option 3) analogously to the recommended test yields a signal.
 
\section{Conclusion}\label{sec-conclusion}

We presented the R package WSPsignal, which implements the family of WSP signal detection tests for right‑censored TTE data. The package provides functions for applying default frequentist and Bayesian tests, as well as a simulation framework to identify the most suitable test specification for a given signal detection task.

Using two synthetic datasets representing different data scenarios, we demonstrated the complete WSP test workflow with annotated code examples. The examples highlight how the recommended test specification can vary depending on factors such as sample size and prior information.

By integrating all WSP test variants into a single, open‑source package, we aim to lower the technical barrier for applied researchers in statistics, pharmacovigilance, and related fields. Its transparent and reproducible implementation facilitates broader adoption, critical review, and further methodological development within the statistical community.

\newpage 

\section{Acknowledge statement}\label{acknowledge-statement}

\section{Disclosure statement}\label{disclosure-statement}

The authors have no conflicts of interest to declare.

\section{Funding}
The authors declare that there is no funding.

\section{Data Availability Statement}\label{data-availability-statement}
The synthetic datasets \code{muscu} and \code{muscu2} are available in the R package WSPsignal.

\phantomsection\label{supplementary-material}
\bigskip

\begin{center}

{\large\bf SUPPLEMENTARY MATERIAL}

\end{center}

\begin{description}
\item[R package WSPsignal:]
R-package WSPsignal (available on GitHub) containing code to perform the WSP signal detection methods
described in the article. The package also contains all datasets used as
examples in the article. 
\item[R scripts:]
R scripts to follow along and replicate the exemplary applications presented in the article (available on \url{osf.io/h7edn}).
\end{description}

\appendix 

\newpage
\section{Computational performance of Bayesian tests}\label{app:computational_perf}

\begin{table}[h]
\centering
\caption{Outcome from \texttt{eval.non\_conv\_cases(pc\_list2)} showing an overview of planned and not-run simulations grouped by WSP model and prior distribution.}\label{tab:demoappl-musku2-notrun}
\footnotesize
\begin{tabular}{|llrrr|}
\hline
TTE dist. & prior dist. & total planned & total not run & proportion not run \\
\hline
dw  & gg & 2800 & 27 & 0.010 \\
dw  & ll & 2800 & 25 & 0.009 \\
pgw & gg & 2800 &  0 & 0       \\pgw & ll & 2800 &  0 & 0       \\
\hline
\end{tabular}
\end{table}

 \begin{figure}[p]
    \centering
    \includegraphics[width = 0.7\textwidth]{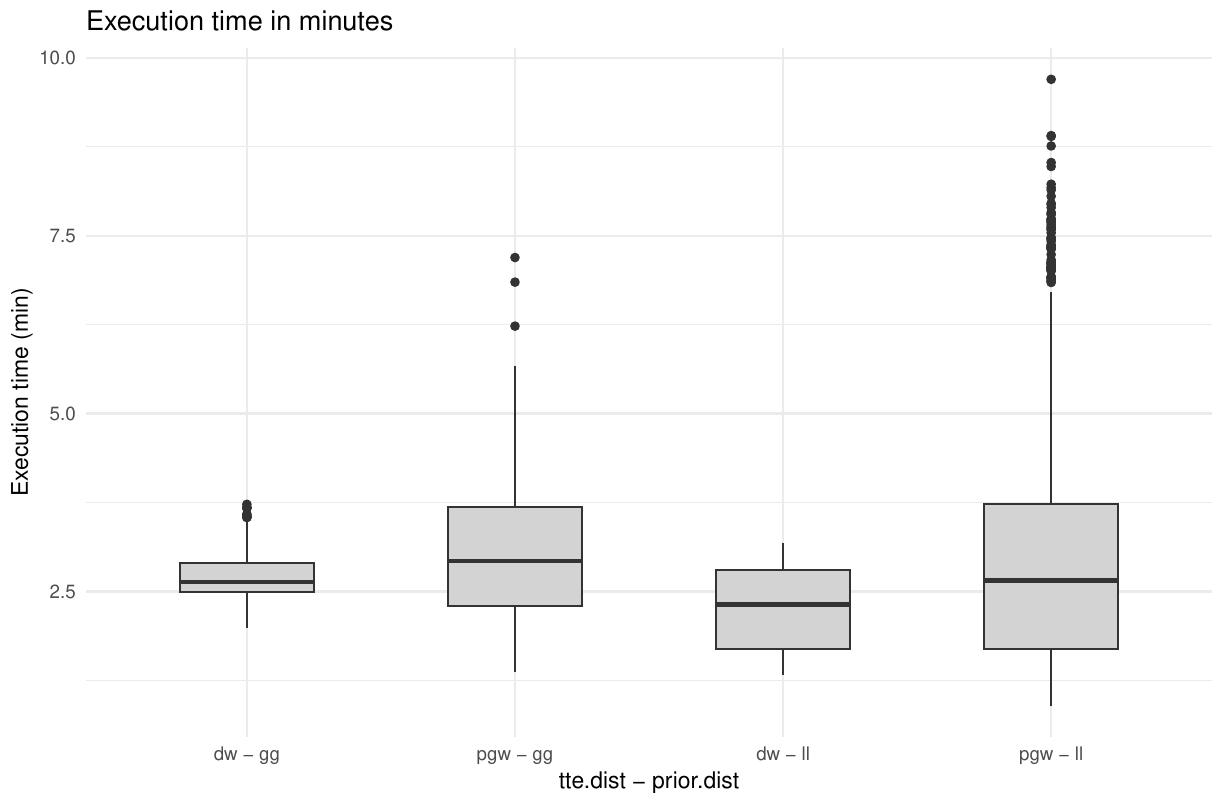}
    \caption{Outcome from \texttt{eval.execution\_times(pc\_list2)}. Boxplots comparing execution times in minutes grouped by WSP model and prior distributions. All execution times are below 10 minutes.}
    \label{fig:demoappl-musku2-runningtimes}
\end{figure}

 \begin{figure}[p]
    \centering
    \includegraphics[width = 0.7\textwidth]{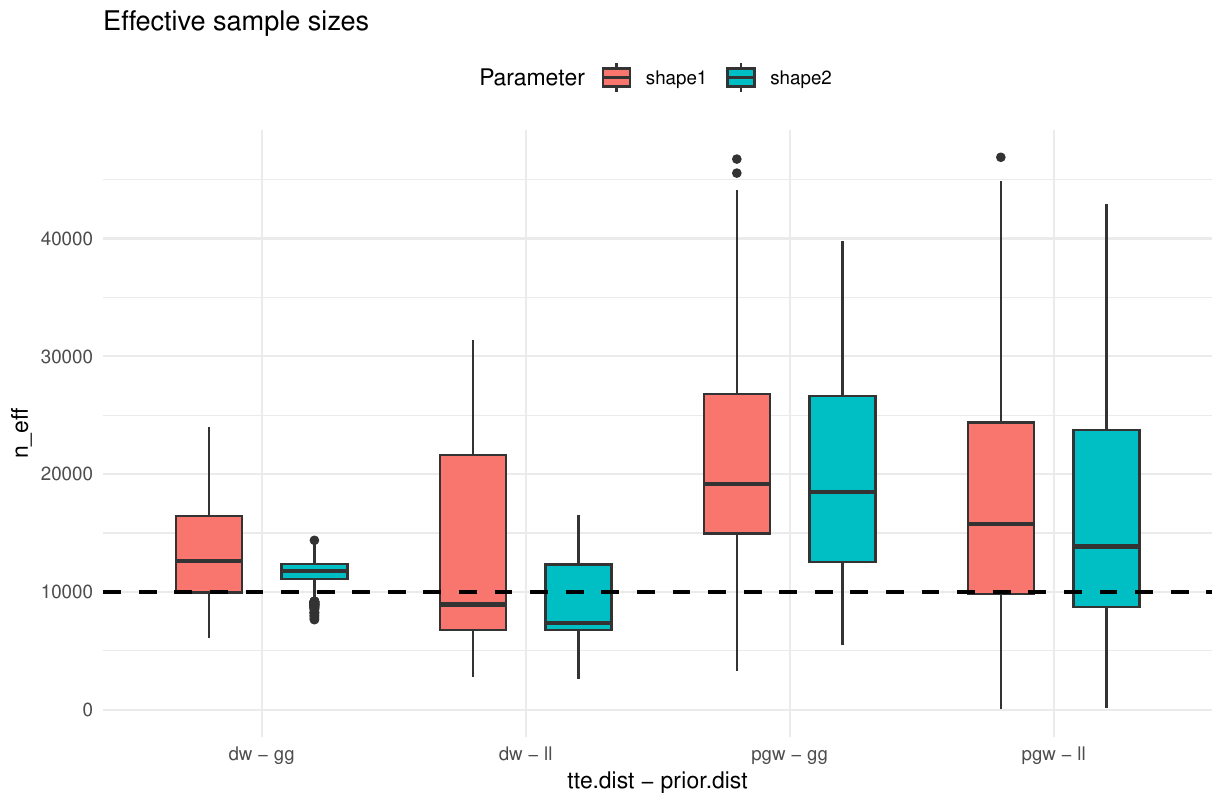}
    \caption{Outcome from \texttt{eval.eff\_sample\_sizes(pc\_list2)}. Boxplots of effective sample sizes (n\_eff) grouped by WSP model and prior distribution for parameters shape/shape1 (red) and shape2/powershape (blue).
    The recommended threshold for the effective sample size is marked with a horizontal dashed line at $10\,000$.
    }
    \label{fig:demoappl-musku2-effectivesamplesize}
\end{figure}

\newpage
\bibliography{bibliography.bib}

@Manual{WSPsignal,
    title = {WSPsignal: Bayesian Weibull Shape Parameter Tests For Signal Detection},
    author = {Julia Dyck and Odile Sauzet},
    year = {2025},
    note = {R package version 1.0.0},
    url  = {http://github.com/julia-dyck/WSPsignal/}
  }

@article{dyck2025bpgw,
      title={The {BPgWSP} test: a {Bayesian} Weibull Shape Parameter signal detection test for adverse drug reactions}, 
      author={Julia Dyck and Odile Sauzet},
      year={2025},
      journal={preprint},
      eprint={2412.05463},
      archivePrefix={arXiv},
      primaryClass={stat.ME},
      note={https://arxiv.org/abs/2412.05463}, 
}

@article{meyboom1997pharmacovigilance,
  title={Principles of signal detection in pharmacovigilance},
  author={Meyboom, RHB and Egberts, ACG and Edwards, IR and Hekster, YA and de Koning, FH and Gribnau, FW},
  journal={Drug safety},
  volume={16},
  pages={355--365},
  year={1997},
  publisher={Springer},
  doi={10.2165/00002018-199716060-00002},
  note={http://dx.doi.org/10.2165/00002018-199716060-00002}
}

@misc{EMAwebsiteADR,
  author = {{European Medicines Agency}},
  title = {Adverse drug reaction},
  howpublished = {https://www.ema.europa.eu/en/glossary-terms/adverse-drug-reaction},
  note = {Accessed: 2024-11-08}
}

@misc{EMAwebsiteAE,
  author = {{European Medicines Agency}},
  title = {Adverse event},
  howpublished = {https://www.ema.europa.eu/en/glossary-terms/adverse-event},
  note = {Accessed: 2025-09-30}
}

@article{zorych2013disproportionality,
  title={Disproportionality methods for pharmacovigilance in longitudinal observational databases},
  author={Zorych, I and Madigan, D and Ryan, P and Bate, A},
  journal={Statistical Methods in Medical Research},
  volume={22},
  number={1},
  pages={39--56},
  year={2013},
  publisher={SAGE Publications Sage UK: London, England},
  doi={10.1177/0962280211403602},
note={http://dx.doi.org/10.1177/0962280211403602}
}

@article{schuemie2011lgps,
  title={Methods for drug safety signal detection in longitudinal observational databases: {LGPS} and {LEOPARD}},
  author={Schuemie, MJ},
  journal={Pharmacoepidemiology and Drug Safety},
  volume={20},
  number={3},
  pages={292--299},
  year={2011},
  publisher={Wiley Online Library},
  doi={10.1002/pds.2051},
note={http://dx.doi.org/10.1002/pds.2051}
}

@article{noren2010temporal,
  title={Temporal pattern discovery in longitudinal electronic patient records},
  author={Nor{\'e}n, GN and Hopstadius, J and Bate, A and Star, K and Edwards, IR},
  journal={Data Mining and Knowledge Discovery},
  volume={20},
  pages={361--387},
  year={2010},
  publisher={Springer},
  doi={10.1007/s10618-009-0152-3},
note={http://dx.doi.org/10.1007/s10618-009-0152-3}
}

@article{hauben2005,
author = {M Hauben and D Madigan and CM Gerrits and L Walsh and EP {Van Puijenbroek}},
title = {The role of data mining in pharmacovigilance},
journal = {Expert Opinion on Drug Safety},
volume = {4},
number = {5},
pages = {929-948},
year  = {2005},
publisher = {Taylor & Francis},
doi = {10.1517/14740338.4.5.929},
note ={https://doi.org/10.1517/14740338.4.5.929}
}

@article{cornelius2012,
  title={A signal detection method to detect adverse drug reactions using a parametric time-to-event model in simulated cohort data},
  author={Cornelius, VR and Sauzet, O and Evans, SJW},
  journal={Drug safety},
  volume={35},
  pages={599--610},
  year={2012},
  publisher={Springer},
note ={https://doi.org/10.2165/11599740-000000000-00000}
}

@Article{sauzet2022,
  author   = {O Sauzet and VR Cornelius},
  journal  = {Frontiers in Pharmacology},
  title    = {Generalised weibull model-based approaches to detect non-constant hazard to signal adverse drug reactions in longitudinal data},
  year     = {2022},
  doi      = {10.3389/fphar.2022.889088},
note ={https://doi.org/10.3389/fphar.2022.889088}
}

@article{Sauzet2024,
  author = {Sauzet, O and Dyck, J and Cornelius, VR},
  doi = {10.1007/s40264-024-01460-2},
  note ={https://doi.org/10.1007/s40264-024-01460-2},
  issn = {1179-1942},
  journal = {Drug Safety},
  number = 11,
  pages = {1149--1156},
  publisher = {Springer Science and Business Media LLC},
  title = {Optimal Significance Levels and Sample Sizes for Signal Detection Methods Based on Non-constant Hazards},
  volume = 47,
  year = 2024
}

@article{exposure_models2024preprint,
      title={An Exposure Model Framework for Signal Detection based on Electronic Healthcare Data}, 
      author={L Dijkstra and T Schink and R Foraita},
      journal = {preprint},
      year={2024},
      eprint={2404.14213},
      archivePrefix={arXiv},
      primaryClass={stat.ME},
      note={https://arxiv.org/abs/2404.14213}
}

@article{coste2023methodreview,
  title={Methods for drug safety signal detection using routinely collected observational electronic health care data: A systematic review},
  author={Coste, A and Wong, A and Bokern, M and Bate, A and Douglas, IJ},
  journal={Pharmacoepidemiology and Drug Safety},
  volume={32},
  number={1},
  pages={28--43},
  year={2023},
  publisher={Wiley Online Library},
  doi={10.1002/pds.5548},
  note ={https://doi.org/10.1002/pds.5548}
}

@book{bagdonavicius2001,
  title={Accelerated Life Models: Modeling and Statistical Analysis},
  author={Bagdonavicius, V and Nikulin, M},
  year={2001},
  publisher={Chapman and Hall/CRC},
  doi={10.1201/9781420035872 },
  note={https://doi.org/10.1201/9781420035872}
}

@article{sauzet2013illustration,
  title={Illustration of the weibull shape parameter signal detection tool using electronic healthcare record data},
  author={Sauzet, O and Carvajal, A and Escudero, A and Molokhia, M and Cornelius, VR},
  journal={Drug safety},
  volume={36},
  number={10},
  pages={995--1006},
  year={2013},
  publisher={Springer},
  doi={10.1007/s40264-013-0061-7},
  note={https://doi.org/10.1007/s40264-013-0061-7}
}

@article{fawcett2004,
  author={T Fawcett},
  title={ROC Graphs: Notes and Practical Considerations for Researchers},
journal = {preprint},
  year={2007},
  note={https://api.semanticscholar.org/CorpusID:2247957},
}

@article{lloyd1998,
  title={Using {S}moothed {R}eceiver {O}perating {C}haracteristic {C}urves to {S}ummarize and {C}ompare {D}iagnostic {S}ystems},
  author={Lloyd, CJ},
  journal={Journal of the American Statistical Association},
  volume={93},
  number={444},
  pages={1356--1364},
  year={1998},
  publisher={Taylor \& Francis},
  doi={10.1080/01621459.1998.10473797},
  note={https://doi.org/10.1080/01621459.1998.10473797}
}

@book{kruschke2015,
  author    = {Kruschke, J},
  publisher = {Academic Press},
  title     = {Doing Bayesian Data Analysis},
  edition = {2nd},
  year      = {2015},
  pages  = {87}
}

@article{kruschke2018,
author = {J Kruschke},
title ={Rejecting or {A}ccepting {P}arameter {V}alues in {B}ayesian {E}stimation},
journal = {Advances in Methods and Practices in Psychological Science},
volume = {1},
number = {2},
pages = {270-280},
year = {2018},
doi = {10.1177/2515245918771304},
note={https://doi.org/10.1177/2515245918771304}
}

@misc{THINwebsite,
  title        = {{THIN}: The Health Improvement Network},
  author       = {{The Health Improvement Network}},
  note         = {Accessed: 2024-09-30},
  howpublished = {https://www.the-health-improvement-network.com/}
}

@article{adr_bis_muscu,
title = "Adverse effects of bisphosphonates: Implications for osteoporosis management",
author = "Kennel, KA and Drake, MT",
year = "2009",
month = jul,
doi = "10.4065/84.7.632",
language = "English (US)",
volume = "84",
pages = "632--638",
journal = "Mayo Clinic proceedings",
issn = "0025-6196",
publisher = "Elsevier Ltd",
number = "7",
note = {https://doi.org/10.1016/S0025-6196(11)60752-0}
}

\end{document}